\documentclass[aps,prl,twocolumn,superscriptaddress,citeautoscript,amssymb,reprint,showpacs,floatfix,longbibliography]{revtex4-1}
\usepackage[ansinew]{inputenc}
\usepackage[T1]{fontenc}
\usepackage{float}
\usepackage{amsmath}
\usepackage{ulem}
\usepackage{mathptmx}
\usepackage{mathrsfs}
\usepackage[thinspace,thinqspace]{SIunits}
\usepackage[english]{babel}
\usepackage{graphicx}
\usepackage{color}
\usepackage{natbib}

\begin{document}
\author{J. F. Landaeta}
\affiliation{Institute of Solid State and Materials Physics, TU Dresden, 01069, Dresden, Germany}
\affiliation{Max Planck Institute for Chemical Physics of Solids, 01187 Dresden, Germany}

\author{K. Semeniuk}
\affiliation{Max Planck Institute for Chemical Physics of Solids, 01187 Dresden, Germany}

\author{J. Aretz}
\affiliation{Max Planck Institute for Chemical Physics of Solids, 01187 Dresden, Germany}

\author{K. Shirer}
\affiliation{Max Planck Institute for Chemical Physics of Solids, 01187 Dresden, Germany}

\author{D. A. Sokolov}
\affiliation{Max Planck Institute for Chemical Physics of Solids, 01187 Dresden, Germany}

\author{N. Kikugawa}
\affiliation{National Institute for Materials Science, Tsukuba, 305-0003, Japan}

\author{Y. Maeno}
\affiliation{5 Toyota Riken - Kyoto University Research Center (TRiKUC), Kyoto 606-8501, Japan}

\author{I. Bonalde}
\affiliation{Centro de F\'isica, Instituto Venezolano de Investigaciones Cient\'ificas, Apartado 20632, Caracas 1020-A, Venezuela}

\author{J. Schmalian}
\affiliation{Institute for Theoretical Condensed Matter Physics, Karlsruhe Institute of Technology, 76131 Karlsruhe, Germany}
\affiliation{Institute for Quantum Materials and Technologies, Karlsruhe Institute of Technology, 76131 Karlsruhe, Germany}

\author{A. P. Mackenzie}
\affiliation{Max Planck Institute for Chemical Physics of Solids, 01187 Dresden, Germany}
\affiliation{Scottish Universities Physics Alliance, School of Physics and Astronomy, University of St. Andrews, St. Andrews KY16 9SS, United Kingdom}
\author{E. Hassinger}
\thanks{correspondence should be addressed to javier.landaeta@cpfs.mpg.de or elena.hassinger@tu-dresden.de}
\affiliation{Institute of Solid State and Materials Physics, TU Dresden, 01069, Dresden, Germany}
\affiliation{Max Planck Institute for Chemical Physics of Solids, 01187 Dresden, Germany}

\title{Evidence for vertical line nodes in Sr$_2$RuO$_4$ from nonlocal electrodynamics}
\date{\today}

\begin{abstract}
By determining the superconducting lower  and upper critical fields $H_\mathrm{c1}(T)$ and $H_\mathrm{c2}(T)$, respectively, in a high-purity spherical Sr$_2$RuO$_4$ sample via ac-susceptibility measurements, we obtain the temperature dependence of the coherence length $\xi$ and the penetration depth $\lambda$ down to 0.04\,$T_c$. Given the high sample quality,  the observed $T^2$ dependence of $\lambda$ at low temperatures  cannot be explained in terms of impurity effects. Instead, we argue that the weak type-II superconductor Sr$_2$RuO$_4$ has to be treated in the non-local limit. In that limit, the  penetration depth data agree with a gap structure having vertical line nodes, while horizontal line nodes cannot account for the observation.  
\end{abstract}
\maketitle
\section{Introduction}
Understanding the superconductivity of Sr$_2$RuO$_4$ is a challenge that has now spanned nearly three decades \cite{Maeno1994}.  The high purity of the crystals available for study, combined with the relatively simple and well-understood normal state \cite{Mackenzie1996} means that this should be a soluble problem, and it has become a milestone for the whole field of unconventional superconductivity \cite{Mackenzie2003, Mackenzie2020, Mackenzie2017, MAENO2012,Kallin2012,Liu2015,Leggett2021}.  Progress is hindered by the lack of a complete understanding of its superconducting order parameter.  Recent studies of the spin susceptibility in the superconducting state have called into question the long-held paradigm of a spin-triplet, odd-parity order parameter, and provided strong evidence for a spin singlet, even-parity state \cite{Pustogow2019,Ishida2020,Chronister2021}.  The question of whether the order parameter breaks time-reversal symmetry or not is also the subject of ongoing investigations \cite{Luke1998,Grinenko2021,Li2021,Willa2021}.  

To further inform the rejuvenated theoretical effort that these new results have stimulated, it is important to find new ways to address one of the other issues about which there is apparently conflicting information: the nodal structure of the superconducting gap.  It is widely agreed, on the basis of, for example, ultrasound \cite{Lupien2001}, penetration depth \cite{Bonalde2000}, heat capacity \cite{Deguchi2004,Kittaka2018,Izawa2001} and thermal conductivity \cite{Tanatar2001,Suzuki2002,Hassinger2017} that the gap in Sr$_2$RuO$_4$ has nodes, but different conclusions have been reached about whether these are horizontal \cite{Kittaka2018,Iida2020} or vertical \cite{Deguchi2004,Tanatar2001,Hassinger2017}.  Any information on this issue is important, because horizontal line nodes would imply mechanisms incorporating inter-plane or inter-orbital pairing \cite{Machida2020,Ramires2019,Suh2020,Palle2023}, while vertical line nodes would be consistent with pairing states formed from in-plane electronic states \cite{Roising2019,Romer2020,Kivelson2020,Palle2023}, the latter being natural for a quasi-2D material.

In this paper, we approach the problem using a technique that has not so far been employed in the study of Sr$_2$RuO$_4$.  We perform magnetic susceptibility measurements on a nearly spherical sample sculpted from an ultra-high purity single crystal with $T_{\rm c} = 1.5\,{\rm K}$ for $\mu_0 H \parallel c$.  The measurements enable a quantitative determination of the temperature dependence of both the lower and upper critical fields, $H_{c1}$ and $H_{c2}$.  From this information, we derive the temperature dependence of the in-plane penetration depth $\lambda$ and the superconducting coherence length $\xi$. From the former, we show that non-local electrodynamics, proposed in Ref.~\cite{Kosztin1997} for nodal superconductors, must be used to analyse the penetration depth results in Sr$_2$RuO$_4$ with this level of purity.  Calculations in this non-local limit distinguish the responses from vertical and horizontal line nodes, and our results are shown to be consistent with the prediction for vertical nodes.

\section{Experimental results}

\begin{figure}[htbp]
	\centering
	\includegraphics[width=\linewidth]{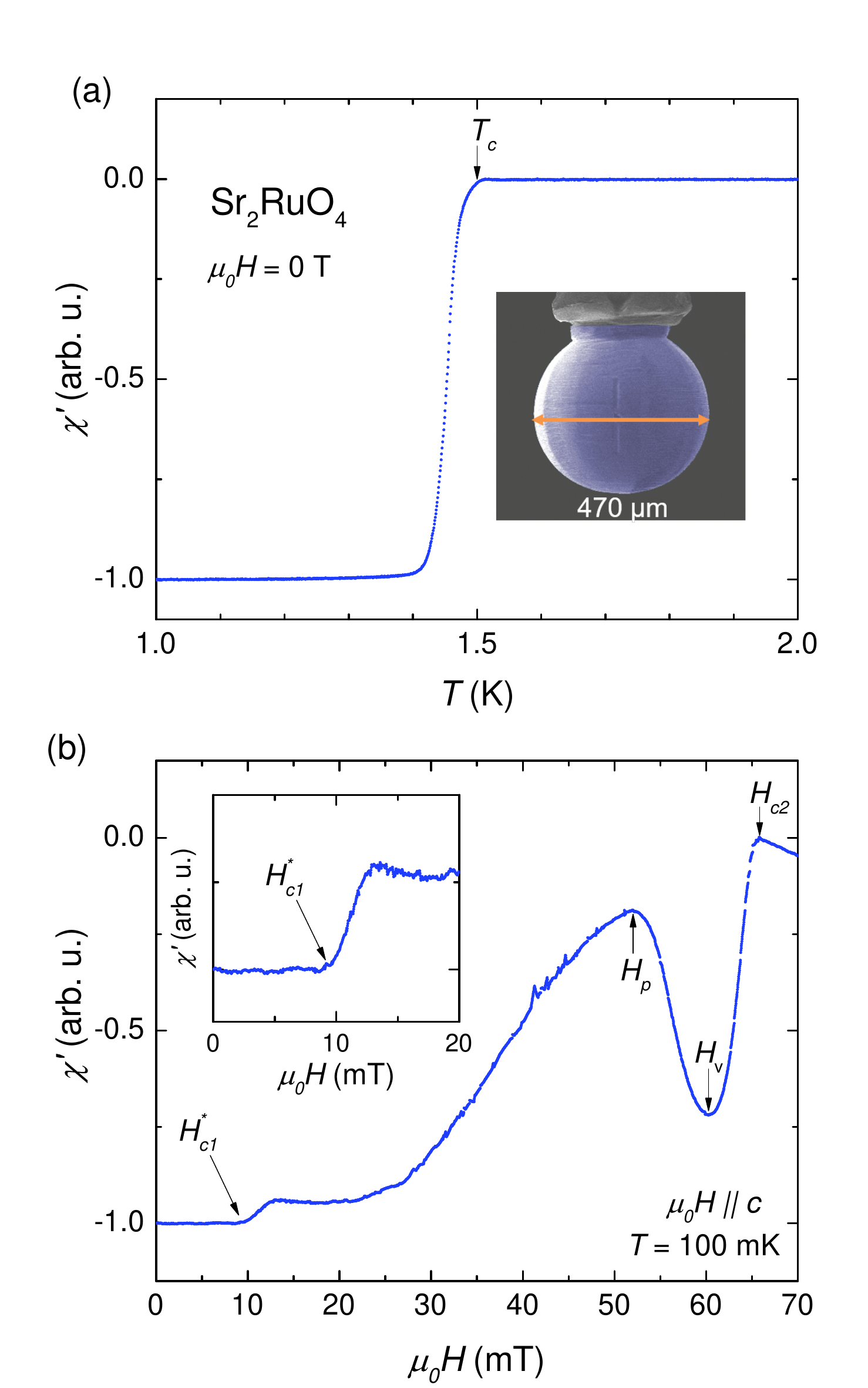}
	\caption{(a) Temperature dependence of ac susceptibility $\chi'(T)$  of the spherical sample of Sr$_2$RuO$_4$. The $T_c$ is defined at the onset of the superconducting transition, highlighted with an arrow. (b) Magnetic field dependence (up sweep) of $\chi'(H)$ at 100 mK. The upper critical field $H_\mathrm{c2}$ is defined at the onset of the transition, and the features related to the vortex physics, $H_\mathrm{p}$ and $H_\mathrm{v}$, are defined at the peak and valley as indicated by arrows. The inset shows a zoom  near $H^*_\mathrm{c1}$, defined as the field where the susceptibility departs from the constant minimum value.}
\label{fig:Figure_1}
\end{figure}

To study the critical fields of Sr$_2$RuO$_4$, we measured the magnetic field dependence of the ac-susceptibility at different temperatures with the external magnetic field $\mu_0H$ applied parallel to the crystallographic c axis. We used a high-purity sample 
cut into a sphere with a diameter of 470 $\mu$m using focussed ion beam (FIB) milling as shown in Fig.~\ref{fig:Figure_1}(a) (for details, see Ref.~\cite{Supplementary2023}). The spherical shape gives a well-defined demagnetization factor independent of the magnetic field direction  and removes uncertainties from other shapes that can strongly influence the measured field values, particularly that of $H_\mathrm{c1}$. 

The temperature dependence of the real part of the magnetic ac-susceptibility $\chi'(T)$ at $H = 0$ is shown in Fig.~\ref{fig:Figure_1}(a). The superconducting critical temperature $T_c=1.5$\,K reveals a high sample purity, comparable to samples having the highest  $T_c$ values \cite{Mackenzie1998}. Figure \ref{fig:Figure_1}(b) shows the field dependence of the ac susceptibility $\chi'(H)$ at $100\,{\rm mK}$. These are the up sweep data that have been corrected for the remanent field of the magnet \cite{Supplementary2023}. We identify four features: the lower critical field $H_\mathrm{c1}^{*}$, which is uncorrected for demagnetisation, the upper critical field $H_\mathrm{c2}$, a peak $H_\mathrm{p}$, and a valley $H_\mathrm{v}$, the latter two being likely associated with the superconducting vortex physics \cite{Yohosida1996,Jerzembeck2022}. $H_\mathrm{c2}$ is defined as the onset of the normal state transition and $H_{c1}^{*}$ as the first deviation of the susceptibility from the full screening in the Meissner state,  as defined in the inset to Fig.\ref{fig:Figure_1}(b) (see also the Supplementary information for more details \cite{Supplementary2023}). To obtain the actual lower critical field 
\begin{equation}
H_{c1}=H_{c1}^{*}(1-N)^{-1},
\label{eq_demag}
\end{equation}
knowledge of the demagnetization factor $N$ is required. Crucial for our analysis is that for a spherical sample  $N=1/3$ \cite{Tinkham1974}.
It is important to note that the feature in the ac susceptibility due to $H_\mathrm{c1}^{*}$ remains the same if we sweep the magnetic field up, starting in zero-field cooled or under field-cooled conditions 
(see the supplementary material \cite{Supplementary2023} ). These results show that we can clearly detect a sharp signature at $H^*_\mathrm{c1}$.


\begin{figure}[htbp]
	\centering
	\includegraphics[width=\linewidth]{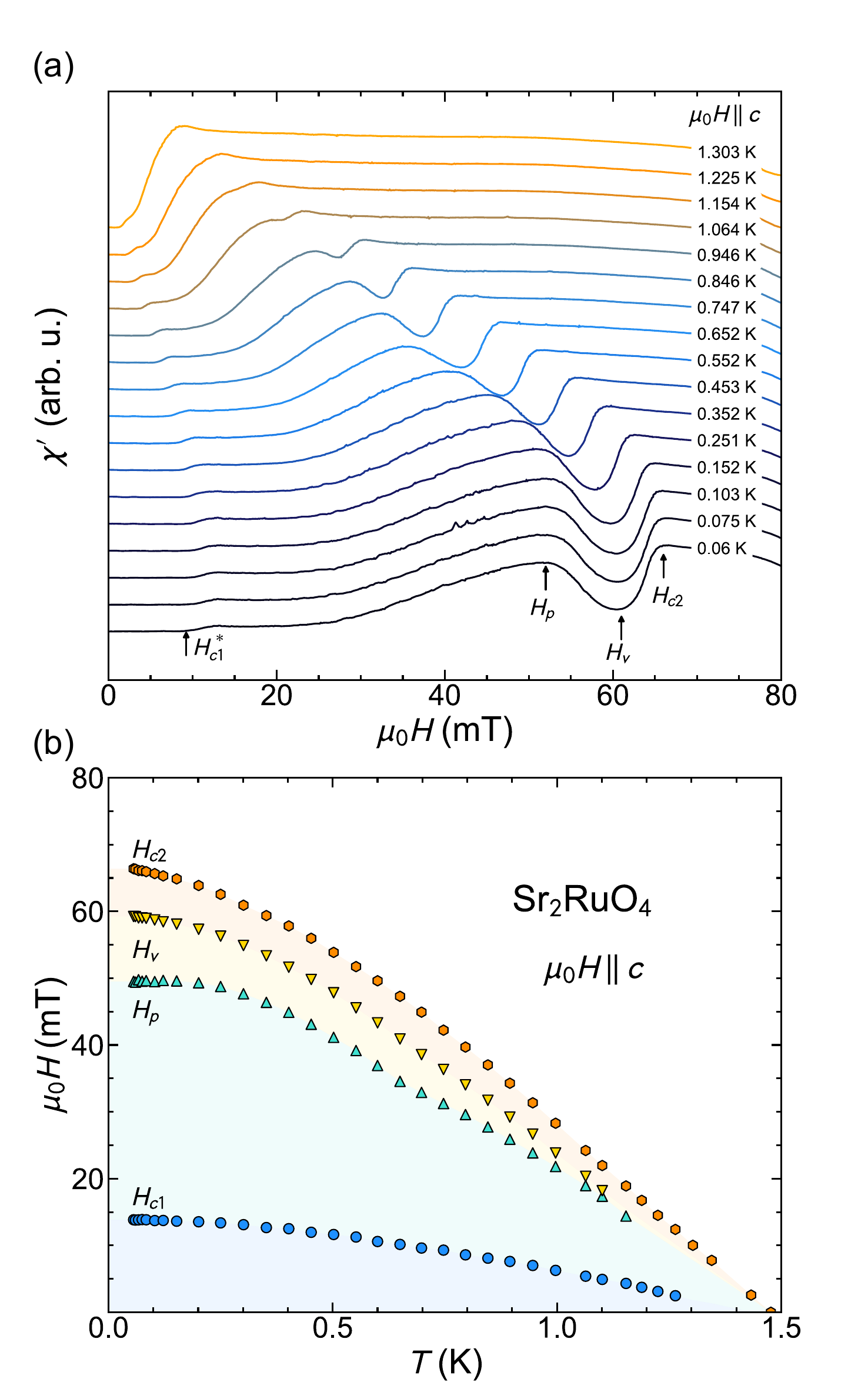}
	\caption{(a) Magnetic field dependence (up sweep) of $\chi'(H)$ at different selected temperatures of the spherical sample with the field applied along the c axis. The $H^*_\mathrm{c1}$, $H_\mathrm{p}$, $H_\mathrm{v}$ and $H_\mathrm{c2}$ are indicated with symbols. (b) Superconducting phase diagram extracted from (a).}
\label{fig:Figure_2}
\end{figure}

Figure \ref{fig:Figure_2} (a) exhibits the magnetic field dependence of the susceptibility at different temperatures. The four features described in Figure \ref{fig:Figure_1}(b) are indicated with arrows on the susceptibility curves.
 From this study, we identified the signature of $H^*_\mathrm{c1}$ up to 1.25\,K and the peak-valley ($H_\mathrm{p}$-$H_\mathrm{v}$) vortex features up to 1.1 K. The $H_\mathrm{p}$ and $H_\mathrm{v}$ features approach each other as we increase the temperature, making them indistinguishable and undetectable for temperatures above 1.1\,K.  

From the data in Fig.~\ref{fig:Figure_2}(b)), we  extrapolated to $T=0$\,K $H^*_\mathrm{c1}(0)=9.27 \, {\rm mT}$, 
$H_\mathrm{p}(0)=49$\,mT, $H_\mathrm{v}(0)=60$\,mT and $H_\mathrm{c2}(0)=67$\,mT. Using the demagnetizing factor of a sphere, we obtain from Eq.~\eqref{eq_demag} for the $T\rightarrow 0$ limit of the lower critical field $H_\mathrm{c1}(0)=13.9\, {\rm mT}$. 

Accurate knowledge of $H_\mathrm{c1}$ is important for the rest of our analysis, so we have checked our results against others from the literature:
 $H^*_\mathrm{c1}=7\, {\rm mT}$ was obtained in specific heat measurements at $60\,{\rm mK}$ in samples with a slab geometry~\cite{Deguchi2004}. A value of $H^*_\mathrm{c1}=7\, {\rm mT}$ was determined using SQUID magnetometry at $T=20\, {\rm mK}$~\cite{Tsuchiya2014}, while  thermal-conductivity  measurements  find $H^*_\mathrm{c1}=8\, {\rm mT}$ at $T=320\,{\rm mK}$~\cite{Tanatar2001} and $12\, {\rm mT}$ for a long plate-shaped sample parallel oriented to the field at $T=300\, {\rm mK}$ \cite{Suzuki2002}.
  Those values reveal a strong influence of the geometry of the sample on the value of $H^*_\mathrm{c1}$. 
For comparison, we use an estimated typical demagnetizing factor $N\approx0$ for a plate-shaped sample oriented parallel to the field while for a slab with proportions of $a\times b \times c=0.5\times 0.5\times 0.33$, $N\approx 0.5$ (see the supplementary material \cite{Supplementary2023}).  This leads to estimated  $H_\mathrm{c1}$ values between 13\,mT and 16\,mT for these measurements. Our value of $H_\mathrm{c1}(0)=13.9\, {\rm mT}$ is hence in very good agreement with  previous measurements using different techniques.

\begin{figure}[htbp]
	\centering
	\includegraphics[width=\linewidth]{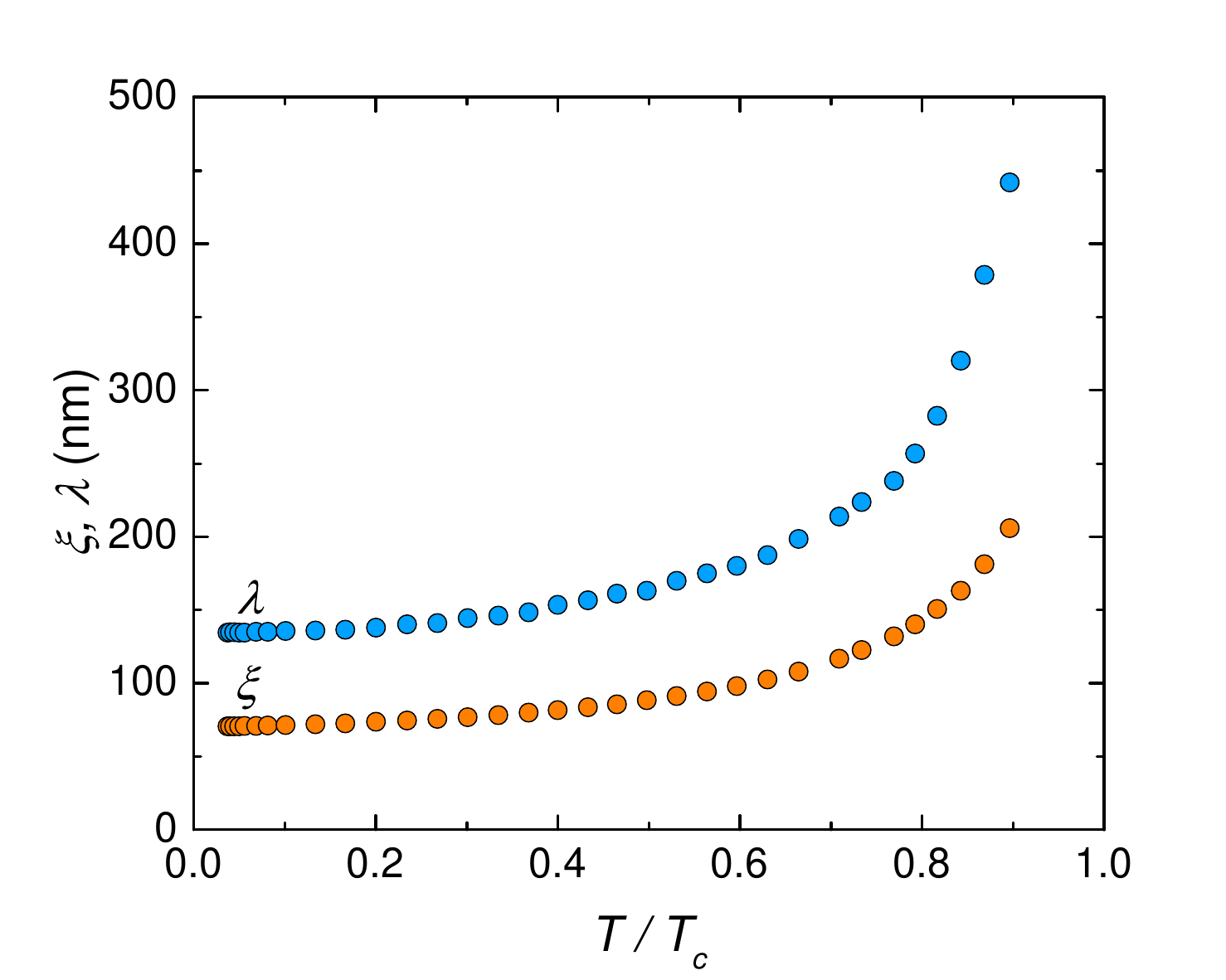}
	\caption{ Temperature dependence of the superconducting coherence length (orange) and the penetration depth (blue) as determined from the upper and lower critical fields via Eq.~\eqref{eq_Hc1c2}. }
\label{fig:Figure_3N}
\end{figure}

We can perform a still more rigorous check of the accuracy of our critical field values by estimating some fundamental superconducting parameters, calculating the thermodynamic critical field, and comparing it to that deduced from specific heat data. To this end, we use 
\begin{eqnarray}
    H_{c2}&= &\frac{\Phi_{0}}{2\pi\xi^{2}} , \nonumber \\
    H_{c1}&=&\frac{\Phi_{0}}{4\pi\lambda^{2}}C\left(\kappa\right),
    \label{eq_Hc1c2}
\end{eqnarray}
that relate the two critical fields with the penetration depth  $\lambda$ and the coherence length $\xi$ of a type II superconductor.  Here, $\Phi_0$ is the flux quantum and  $\kappa=\lambda/\xi>\kappa_c =\tfrac{1}{\sqrt{2}}$ the Ginzburg-Landau parameter. The function $C(\kappa)$ was determined numerically from the solution of the Ginzburg-Landau equations by Brandt \cite{Brandt2003}, who also gave a simple analytic interpolation formula that is highly accurate ($<10^{-3}$) for all values $\kappa >\kappa_c$ and reproduces the limits $C\left(\kappa_c \right)=1$ and $C\left(\kappa \gg 1\right)=\log\kappa+0.49693$~\cite{Hu1972} (see supplementary material \cite{Supplementary2023}).

Using Eq.~\eqref{eq_Hc1c2} and our result for $H_\mathrm{c2}(0)$ and $H_\mathrm{c1}(0)$  yields  $\xi_0=70\, {\rm nm}$ and  $\lambda_0=134\, {\rm nm}$ for the zero-temperature values of coherence length and penetration depth, respectively. This is consistent with the measurement by Muon Spin Rotation ($\mu$SR) $\lambda_0=126$ nm and the calculation of the contribution of each band to the magnetic penetration depth based on Angle-Resolved Photo-electron Spectroscopy (ARPES) $\lambda_0=130$ nm \cite{khasanov2023}.
From $\lambda_0$ and $\xi_0$ we obtain $\kappa_0\equiv\kappa(T=0)=1.92$, i.e. Sr$_2$RuO$_4$ is not a strong type II superconductor as the value of the Ginzburg-Landau parameter is not far from the limit to type I superconductivity ($\kappa_c \approx 0.71$). 
These results yield the thermodynamic critical field $H_c(0)=\frac{H_\mathrm{c2}(0)}{\sqrt{2}\kappa_0}=24.67\,{\rm  mT}$, in very good  agreement with  $H_c=(23\pm2)\,{\rm mT}$ deduced  from specific heat data \cite{Supplementary2023}. In summary, our measurements give precise values of $H_\mathrm{c1}$ and superconducting parameters in full agreement with previous results.


\section{Temperature-dependent Penetration depth  and coherence length}

\begin{figure*}[htbp]
	\centering
	\includegraphics[width=\linewidth]{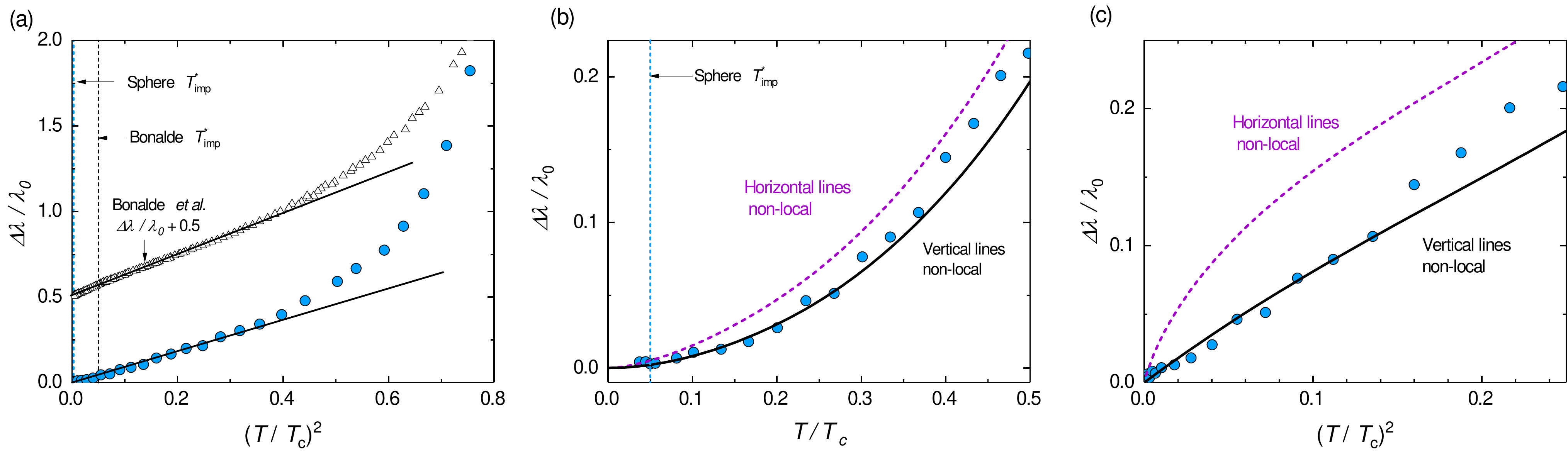}
	\caption{(a) Temperature dependence of $\Delta\lambda/\lambda_0$ vs. $(T/T_c)^2$. The data from Bonalde \textit{et al.}\cite{Bonalde2000} is shifted for clarity. The black and blue dashed lines represent the $T^*_{\mathrm{imp}}$ from Bonalde \textit{et al.} and the spherical sample.(b) Shows the expected behavior of $\Delta\lambda/\lambda_0$ for vertical (solid line) and horizontal (dashed) line nodes in the nonlocal electrodynamics limit.
 (c) Shows $\Delta\lambda/\lambda_0$ vs. $(T/T_c)^2$; in this scale, the difference between the vertical and horizontal nodes becomes clearer in this scale.}
\label{fig:Figure_3}
\end{figure*}

Through Eq.~\eqref{eq_Hc1c2}, the temperature dependencies  of $H_\mathrm{c1}(T)$ and $H_\mathrm{c2}(T)$ are  directly related to the $T$-variation of the in-plane penetration depth $\lambda(T)$ and coherence length $\xi(T)$, respectively. In Fig. 2 we show both length scales as they vary with temperature. As expected, $\lambda(T)$ and $\xi(T)$ grow with temperature. Focussing on the temperature variation of the penetration depth, we show in Figure~\ref{fig:Figure_3}(a))
\begin{equation}
    \frac{\Delta \lambda(T)}{\lambda_0}=\frac{\lambda(T)-\lambda_0}{\lambda_0},
\end{equation}
which exhibits a  $T^2$ dependence for temperatures  below $ \approx 0.5 T_{{\rm c}}$ .

One possible explanation of the $T^2$ dependence,  that is appropriate for systems such as the cuprates is the effect of impurities in  superconductors with line nodes. Hirschfeld and Goldenfeld showed that for an unconventional superconductor with vertical line nodes, the scattering due to impurities would lead to a change in the temperature dependence of $\lambda(T)\propto T$ to $T^2$ below a crossover temperature $T^*_{\mathrm{imp}}\approx 0.83(\Gamma\Delta_{0})^{1/2}$, where $\Gamma$ is the scattering rate and $\Delta_0$ is the magnitude of the superconducting gap\cite{Hirschfeld1993}. 
In our extremely clean sample with $T_{\rm c} \approx 1.5\, {\rm K}$,  we obtain $T^*_{\mathrm{imp}}\leq 0.05T_c$, such that the effect of impurities cannot explain our experiment across the wide range of temperatures where we see the $T^2$ dependence.  

There have been several previous reports of $T^2$ behaviour of $\Delta \lambda/\lambda_0$ in Sr$_2$RuO$_4$ within the Meissner state \cite{Bonalde2000,Baker2009,Ormeno2006}. However,  these were on crystals with lower $T_c$ and higher $T^*_{\mathrm{imp}}$.  Measurements of the penetration depth with tunnel diode oscillator (TDO) technique~\cite{Bonalde2000} are shown in Fig.~\ref{fig:Figure_3}(b) \cite{Bonalde2000}, while Refs.~\cite{Baker2009,Ormeno2006} used
microwave surface impedance measurements. For the samples used for these measurements, we estimate, using the  relationship between the scattering rate $\Gamma$ and $T_c$ \cite{Suzuki2002b},  $T^*_{\mathrm{imp}}\approx 0.2T_c$.  It was therefore uncertain whether the observed behaviour should be attributed to impurity scattering or not.  In our data, there is no such ambiguity: the data shown in Fig.\ref{fig:Figure_3} are in the clean limit.  At first sight, this presents a puzzle.
In the local electrodynamics, applicable to most unconventional superconductors, only point nodes can give a $T^2$ dependence of the penetration depth~\cite{Einzel1986}.  In contrast, all the even parity states under discussion as potential order parameters for Sr$_2$RuO$_4$  either have line nodes or are fully gapped, giving only $T$-linear or exponential dependencies of $\Delta \lambda/\lambda_0$ at low temperatures.

The resolution to this apparent paradox lies in considering non-local electrodynamics. Nonlocal effects in the electromagnetic response below $T_{c}$ go
back to the analysis of Pippard for type I superconductors \cite{Pippard1953}, where spatial modes of the magnetic field with wavelengths smaller than the coherence length give rise to reduced screening currents that shield the external field, affecting the length scale up to which the field can penetrate.

It was pointed out by Kosztin and Leggett \cite{Kosztin1997} that
a nonlocal response can also play a role in type II superconductors
if they possess nodes of the gap function. Qualitatively, this is
due to the effective coherence length $\xi_{\boldsymbol{k}}\sim v_{F}/\Delta_{\boldsymbol{k}}$
that diverges along the nodal directions where the gap function $\Delta_{\boldsymbol{k}}$
vanishes. However, one must keep in mind that in the ground state,
the entire Fermi surface contributes to the phase stiffness and the
relative importance of the nodal points is negligible. On the other
hand, the temperature dependence of the penetration depth is dominated
by thermal quasiparticle excitations near the nodes. The effective
coherence length of those excitations is the thermal de Broglie length
$\xi_{T}\sim v_{F}/T$. The result by Kosztin and Leggett for a system
with vertical line nodes can then be written
in the form: 
\begin{equation}
\frac{\Delta\lambda\left(T\right)}{\lambda_{0}}\sim\left.\frac{\Delta\lambda\left(T\right)}{\lambda_{0}}\right|_{{\rm loc}}\frac{\lambda_{0}}{\xi_{T}},\label{eq:KLqual}
\end{equation}
where $\left.\Delta\lambda\left(T\right)/\lambda_{0}\right|_{{\rm loc}}=\log2\frac{T}{\Delta_{0}}$ is the well-established result
for the local electromagnetic response of a nodal superconductor.  Hence, it follows that
in the non-local limit $\Delta\lambda\left(T\right)/\lambda_{0}\propto \kappa_0 T^2/\Delta_{0}^{2}$.
Here $\Delta_{0}$ is the gap amplitude that also enters in the low-temperature density of states $\rho\left(\omega\right)=\rho_{F} \left|\omega\right|/ \Delta_{0}$ of the nodel superconductor. This
non-locality is tied to the condition $\lambda_{0}\ll\xi_{T}$, which
translates to $T\ll T^{*}=\Delta_{0}/\kappa_0$.
The change in the electromagnetic response due to thermally excited nodal
quasiparticles is reduced for wavelength of the magnetic field smaller
than $\xi_{T}$. In strongly type II superconductors with high $\kappa$, the range of temperatures over which non-local effects are relevant becomes vanishingly small.  However, the low $\kappa$ of Sr$_2$RuO$_4$ leads to a much larger range of temperatures for which the non-local physics is applicable; in the Supplementary Material, we estimate $T^*$ to be as high as $0.8 T_{\rm c}$.

Intriguingly, the non-local regime offers a qualitative distinction between the effects of vertical and horizontal line nodes on $\Delta \lambda/\lambda_0$, because horizontal nodes lie  in the plane of the screening
super-currents. This regime was analyzed by Kusunose and Sigrist in
Ref.\cite{Kusunose2002} and their result for horizontal line nodes can be formulated as 
\begin{equation}
\frac{\Delta\lambda\left(T\right)}{\lambda_{0}}\sim\left.\frac{\Delta\lambda\left(T\right)}{\lambda_{0}}\right|_{{\rm loc}}\frac{\lambda_{0}}{\xi_{T}}\log\frac{\xi_{T}}{\lambda_{0}},\label{eq:KSqual}
\end{equation}
Hence, the suppression of the electromagnetic response by quasiparticles
with horizontal nodes is less strong and, with $\Delta\lambda\left(T\right)/\lambda_{0}\propto T^{2}\log\left(T^{*}/T\right)$,
in principle distinguishable from those of vertical line nodes where
$\Delta\lambda\left(T\right)/\lambda_{0}\propto T^{2}$. While these
qualitative arguments are limited to the regime of lowest temperatures,
in the supplementary material \cite{Supplementary2023}, we demonstrate the full analysis
of the electromagnetic response for horizontal and vertical line nodes.
The theoretical curves,  shown in Figure \ref{fig:Figure_3}(b) alongside the experimental data,
 depend on the two dimensionless
numbers $\kappa_{0}$ that we determined earlier and $2\Delta_{0}/\left(k_{B}T_{c}\right)$. For the theoretical curves shown in the Figure \ref{fig:Figure_3}(c), we used $2\Delta_{0}/\left(k_{B}T_{c}\right)=3.53$ and  $\kappa_0=1.92$. 
 In the supplementary material, we also show data for $2\Delta_{0}/\left(k_{B}T_{c}\right)=3.16$, deduced from the heat capacity dependence $C_s(T)=\alpha T^2$ in the superconducting state \cite{NishiZaki2000}, that allows for very similar conclusions.  

In Fig.~\ref{fig:Figure_3}(c), we replot the same theoretical results and experimental data as a function of $(T/T_c)^2$, which  highlights the qualitative difference in the predictions for horizontal and vertical line nodes at low temperatures.  In both Figs.~\ref{fig:Figure_3} (b) and (c), the predictions for vertical line nodes are a systematically better match to the data than those for horizontal line nodes. This remains true if we allow $2\Delta_{0}/\left(k_{B}T_{c}\right)$ as an open fit parameter.
Our conclusions are robust as long as there are no strong variations of the gap amplitude among the various Fermi surface sheets. 
Then, the data are compatible with the existence of purely vertical line nodes but not compatible with order parameters containing solely horizontal nodes. For mixed order parameters with both types of node, explicit calculations of $\Delta \lambda/\lambda_0$ would be required to determine whether or not the predictions are compatible within experimental error with our data.

In conclusion, we have used measurements of $H_{\rm c1}$ and $H_{\rm c2}$ on an extremely high purity single crystal of Sr$_2$RuO$_4$ to show that its in-plane low-temperature coherence length $\xi_0=70\, {\rm nm}$ and penetration depth $\lambda_0 = 134\, {\rm nm}$, and that $\Delta \lambda/\lambda$ varies as $T^2$ in the clean limit.  Analysis of our results using non-local electrodynamics confirms that the observations are compatible with vertical line nodes in its superconducting order parameter.  Our measurements and analysis are of relevance to the ongoing quest to understand the order parameter symmetry of Sr$_2$RuO$_4$, and invite careful measurement and analysis of $\Delta \lambda/\lambda$ in other unconventional superconductors in which the non-local regime is experimentally accessible.

\textit{Note added.} We note that Ref. \cite{Mueller2023}  reports measurements of $\Delta\lambda$ as a function of uniaxial pressure Sr$_2$RuO$_4$ that also highlight the importance of non-local effects.

\section{Acknowledgements}
We thank D. Bonn, J.-P. Brison, S. Brown, P. Hirschfeld, G. Palle, R. Prozorov, A. Ramires, and  H. Suderow for stimulating exchange and helpful discussions. E.H. acknowledges funding from Deutsche Forschungsgemeinschaft (DFG) via the CRC 1143-247310070 (project C10) and the Wuerzburg-Dresden cluster of excellence EXC 2147 ct.qmat Complexity and Topology in Quantum Matter - project number 390858490. J.S. was supported by the DFG through TRR 288-422213477 Elasto-Q-Mat (project A07). This work is supported by JSPS KAKENHI (No. JP18K04715, No. JP21H01033, and No. JP22K19093).
J.S. also acknowledges the hospitality of KITP, where part of the
work was done. KITP is supported in part by the National Science Foundation under Grant No. NSF PHY1748958 and NSF PHY-2309135.

\setcounter{equation}{0}
\setcounter{figure}{0}
\setcounter{table}{0}
\makeatletter
\renewcommand{\theequation}{S\arabic{equation}}
\renewcommand{\thefigure}{S\arabic{figure}}

\section{Supplementary material}
\subsection{Experimental methods}
The magnetic AC susceptibility was measured using a custom-made pair of compensated pickup coils, each having a length of 2 mm and 4500 turns \cite{Landaeta2022a}. As a drive coil, we used a superconducting modulation coil with an excitation field of 175 $\mu$T at a frequency of 5 Hz. The output signal from the pickup coils was amplified by a low-temperature transformer with a 1:100 amplification ratio, followed by a low-noise amplifier SR560 from Stanford Research Systems. Our experimental setup incorporated a National Instruments 24-bit PXIe-4463 signal generator and 24-bit PXIe-4492 oscilloscope for data acquisition, utilizing digital lock-in amplification. The noise level of our measurements was approximately 70 pV/$\sqrt{Hz}$.

The ac-susceptibility measurements were conducted within an MX400 Oxford dilution refrigerator down to 40 mK. The temperature was measured with a thermometer directly coupled with a conductive high-purity annealed silver wire to the sample, but situated in the compensated region of the magnet. We only show the real part of the magnetic susceptibility since the imaginary part is not relevant here. We corrected for the remnant field of the superconducting magnet, which remained at a constant value within a range of $\pm$2 mT while sweeping the magnetic field between -120 and 120 mT, as carried out in our study. This is explained in detail below. The earth magnetic field is not shielded in our measurements.

\subsection{Sample preparation}

\begin{figure}
	\centering
	\includegraphics[width=\linewidth]{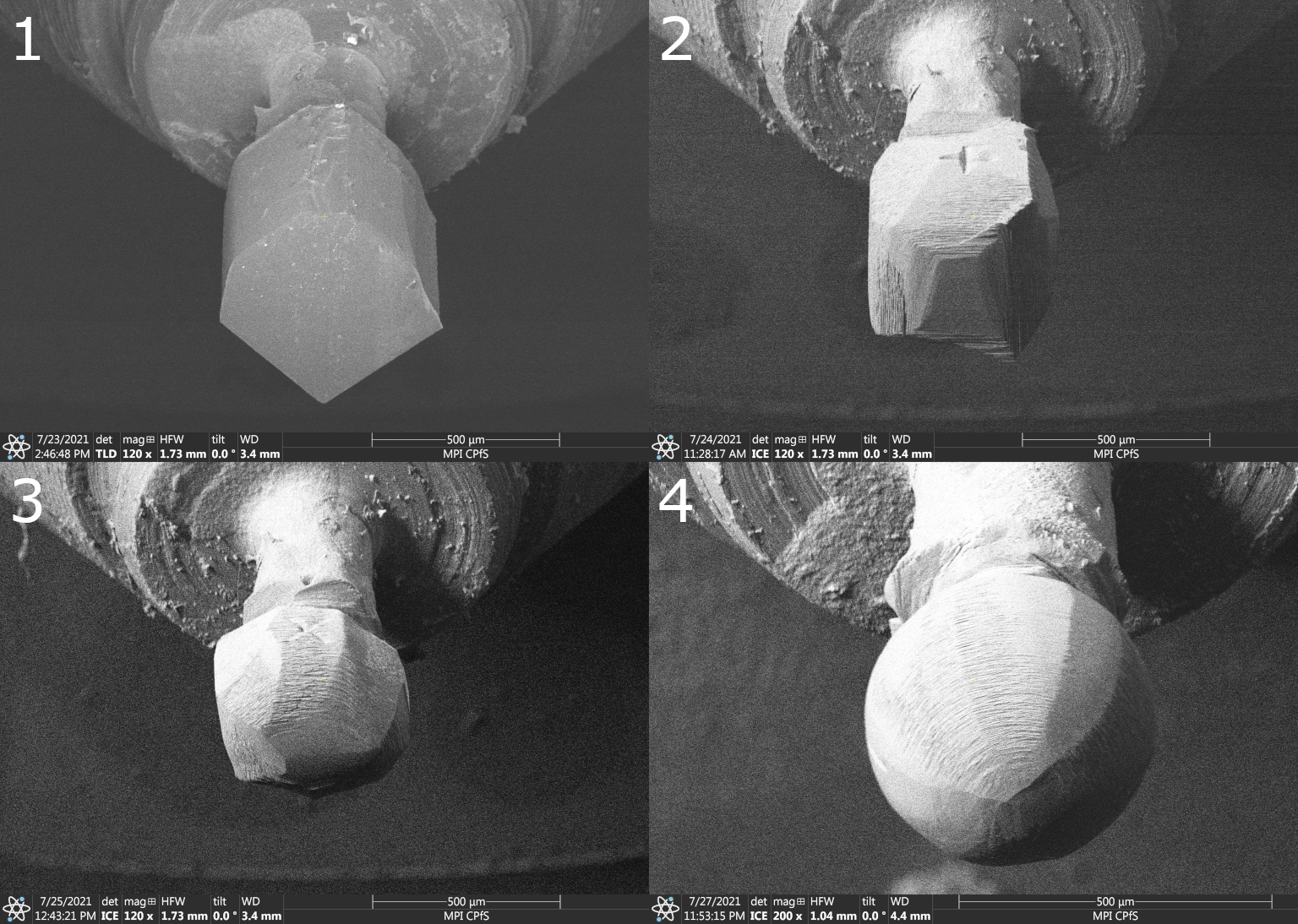}
	\caption{Various stages of fabrication of the spherical Sr$_2$RuO$_4$ sample. The panels are arranged chronologically. The images were taken inside the FIB microscope chamber, using an electron column positioned at 52\degree\ with respect to the ion column.}
\label{fig:fib_sphere}
\end{figure}

The spherical sample of Sr$_2$RuO$_4$ was produced via focused ion beam (FIB) milling. The procedure is illustrated in Fig.~\ref{fig:fib_sphere}. The initial cube-shaped crystal with a side length of 0.5\,mm was mounted on the tip of a rotor of a piezo-actuated stepper motor, which was then mounted onto the specimen platform of a Helios G4 Xe Plasma FIB microscope (manufactured by FEI). The crystal was oriented with its $c$ axis parallel to the motor rotation axis, which in turn was perpendicular to the ion beam. Using the motor, the sample was exposed to the ion beam from different directions, and excess material was ablated by Xe plasma with a voltage of 30\,kV, eventually giving the sample a spherical shape. This energy gives an amorphous surface layer of a thickness of typically 100\,nm. The sample was kept stationary during the milling. The motor was actuated using a controller located outside the microscope chamber. Currents of the order of 1\,$\micro$A were used for the initial pass. These were eventually reduced to the level of several nA as the final shape was approached. To keep track of the $a$ axis direction, a shallow mark was etched on the side of the sphere using the ion beam.
\begin{figure}[htbp]
	\centering
	\includegraphics[width=\linewidth]{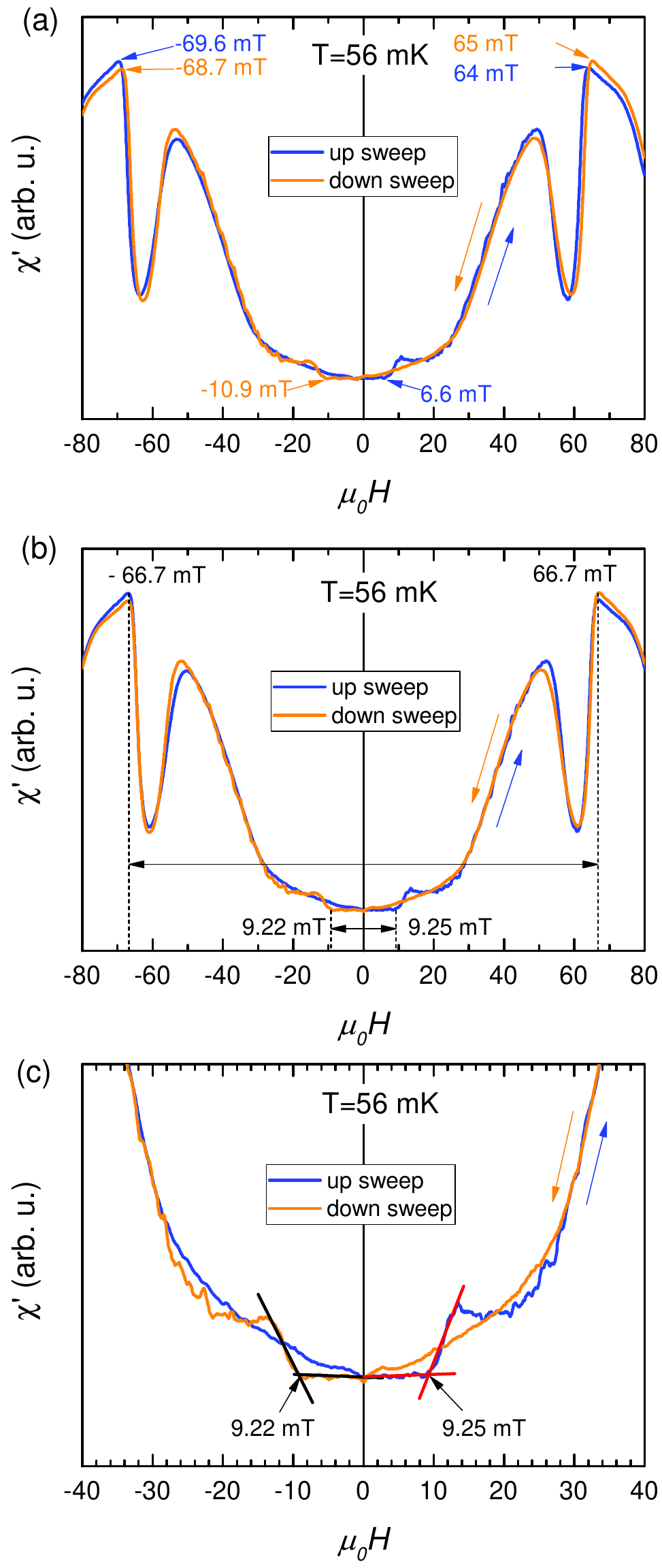}
	\caption{The magnetic field dependence of $\chi'(H)$ of the spherical sample at 56 mK. (a) Raw data of $\chi'(H)$. (b) $\chi'(H)$ with correction in $\mu_0H$ and (c) is a zoom-in of the panel (b) around the magnetic field range of $H_{c1}$. The interception of the lines in (c) corresponds to our definition of $H^*_{c1}$ }
\label{Remanent_field}
\end{figure}
\section{Determination of critical field values}
This paragraph describes the data-analysis procedure used to extract the critical field values necessary for determining the values of penetration depth and coherence length. We carried out measurements of the magnetic susceptibility as a function of temperature for $H=0$ to determine $T_c$, and as a function of the magnetic field at different temperatures. Here, we only show the  real part of the magnetic susceptibility since the imaginary part is not relevant. For each temperature, we start measurements at a magnetic field with an absolute value $\mu_0H> 100$\,mT in the normal state, sweep to the same field of opposite sign and back to the starting field. As will be shown below, these measurements give the same result for $H_{c1}^*$ as starting in the zero-field-cooled condition.
Fig. \ref{Remanent_field}(a) shows the raw data of the real part of the magnetic susceptibility for an example temperature. The curves are shifted along the field axis in a way that there is a shift of both up and down-sweep curves with respect to each other, and that the $H_{c2}$ values (defined here at the maxima indicated by an arrow) are not symmetric with respect to the $H=0$ reading of the magnet. We ascribe this to two effects:
\begin{itemize}
\item[i)]{The up-sweep and down-sweep curves are shifted relative to each other along the field-axis. This shift is of the order of 1\,mT and likely due to a lag during the sweep. This typically occurs in sweep-mode measurements due to the data acquisition loop in which the magnetic field at the time of susceptibility measurement is different from the magnetic field at the time of the field reading, the latter being the recorded and plotted field value.}
\item[ii)]{The asymmetry with respect to the recorded $H=0$ value is of the order 2\,mT and it originates in a remanent field in the superconducting magnet, i.e. trapped flux lines from previous magnetization of the magnet, that leads to a finite remanent field in the center of the magnet even for zero applied current through the superconducting magnet coil. In the example curve, this means that a current corresponding to a field of $\approx -2$\,mT has to be applied so that this remanent field is compensated and that the real magnetic field at the sample is zero.}
\end{itemize}

In order to correct for both effects, we use the fact that $H_\mathrm{c2}$ is expected to be symmetric in field. We read off the value of $H_\mathrm{c2}$ as the maximum in $\chi'$ for each sweep at positive and negative fields -- as indicated by the arrows near $\pm 65$\,mT -- and shift the curve along the field axis so that $H_\mathrm{c2}^{upsweep} = -H_\mathrm{c2}^{downsweep}$.  The result is shown in Fig.~\ref{Remanent_field}b). We use the average of $H_{c2}$ from up and down sweeps to determine the value of $H_{c2}$ used for data analysis. Hysteresis effects can still be observed, especially near $H_{c1}^*$, probably due to vortex pinning. For fields sweeping from the Meissner state up to the vortex state, $H_{c1}^*$ is clearly identified as the field where the susceptibility starts increasing from a constant value, which we determined by construction with two straight lines as shown in Fig. \ref{Remanent_field}(c). When the field is lowered in absolute value so that the Meissner state is entered coming from the vortex state, vortices are slightly pinned and leave the sample slowly. However, at $H=0$ the constant value of the susceptibility is reached and we infer that any pinning effect ends there.

\begin{figure}[htbp]
	\centering
	\includegraphics[width=\linewidth]{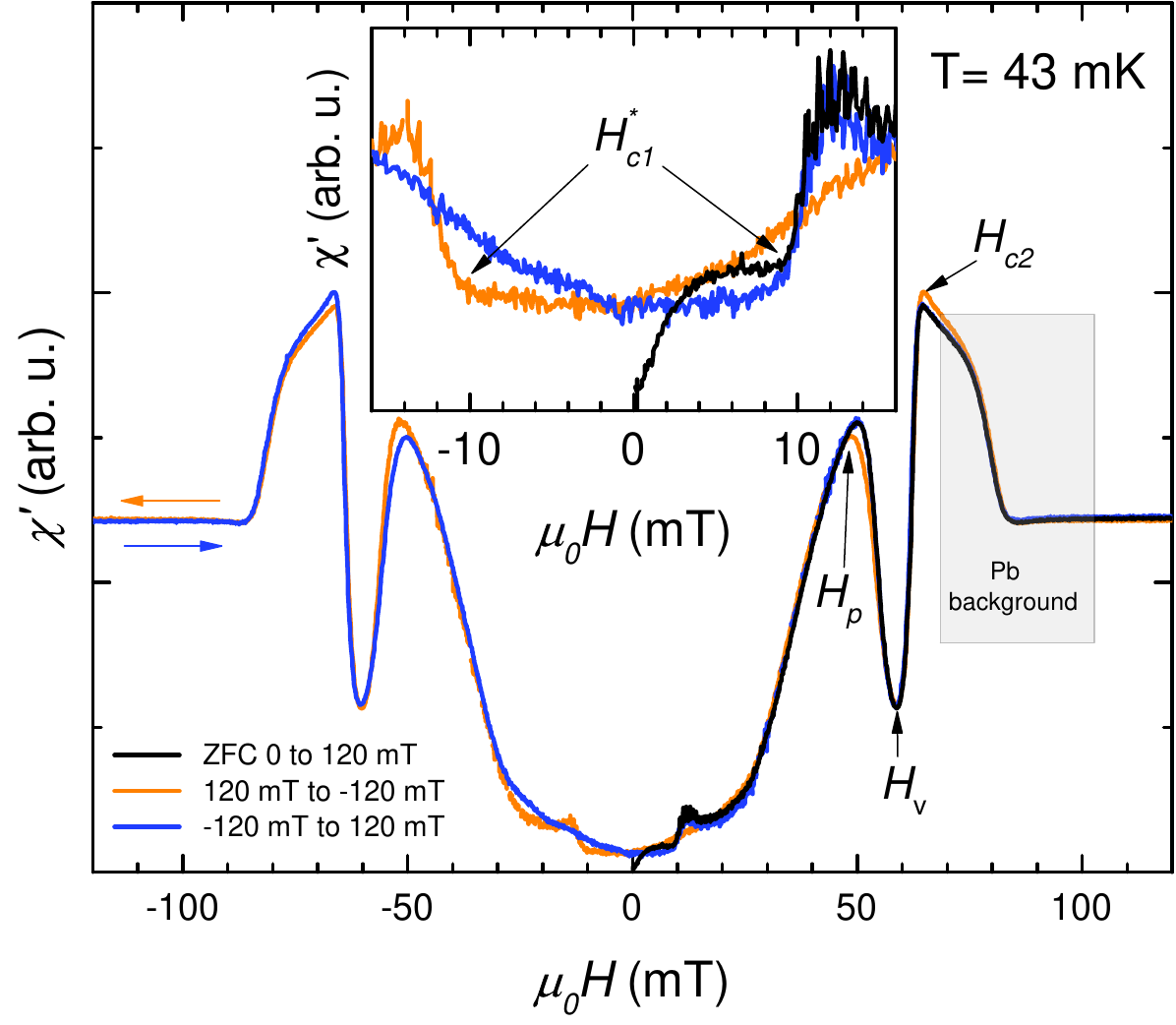}
	\caption{Ac magnetic susceptibility measurements at 43mK. The black line show the measurement  measures with zero field cooling (ZFC) 0 mT to 120 mT, starting with the fully demagnetized superconducting magnet. Blue and orange lines are the subsequent measurements from 120 mT to -120 mT (blue) and -120 mT to 120 mT (orange). The inset shows a zoom-in in the region where $H^*_{c1}$ is defined. The shaded region is the background signal from the Pd of the solder joint and the shield of the low temperature transformer.}
\label{fig:Figure_s3_zfc}
\end{figure}

To exclude any effect of vortex pinning on the value of $H_{c1}^*$ in our measurement procedure, we also compared the results to a zero-field-cooled measurement for 42\,mK in fig. \ref{fig:Figure_s3_zfc}. Here, the whole dilution refrigerator and superconducting magnet were warm before starting the measurement so that we can exclude any effect of remanent fields. The upturn at the beginning of the sweep is then likely related to the voltage induced by the imperfect balance pickup coils when the background changes with small fields by a magnetisation of surrounding parts of the cryostat upon first application of the magnetic field. Apart from this difference below 4\,mT, the same signature near $H_{c1}^*$ is observed as in the measurements coming for high fields. The real part of the magnetic susceptibility is flat below $H_{c1}^*$ and then starts to increase quite sharply. Hence, we are confident that flux pinning does not affect our value of $H_{c1}$ in the measurements where we start in the field-induced normal state. Note that, since $\lambda_0=134$\,nm is of the size of the sample roughness given by the thickness of the amorphous layer created by the ion beam of the order of 100\,nm, the Bean-Levingston surface barrier effect \cite{Liang2005} should not play a role here.

\begin{figure}[t]
	\centering
	\includegraphics[width=\linewidth]{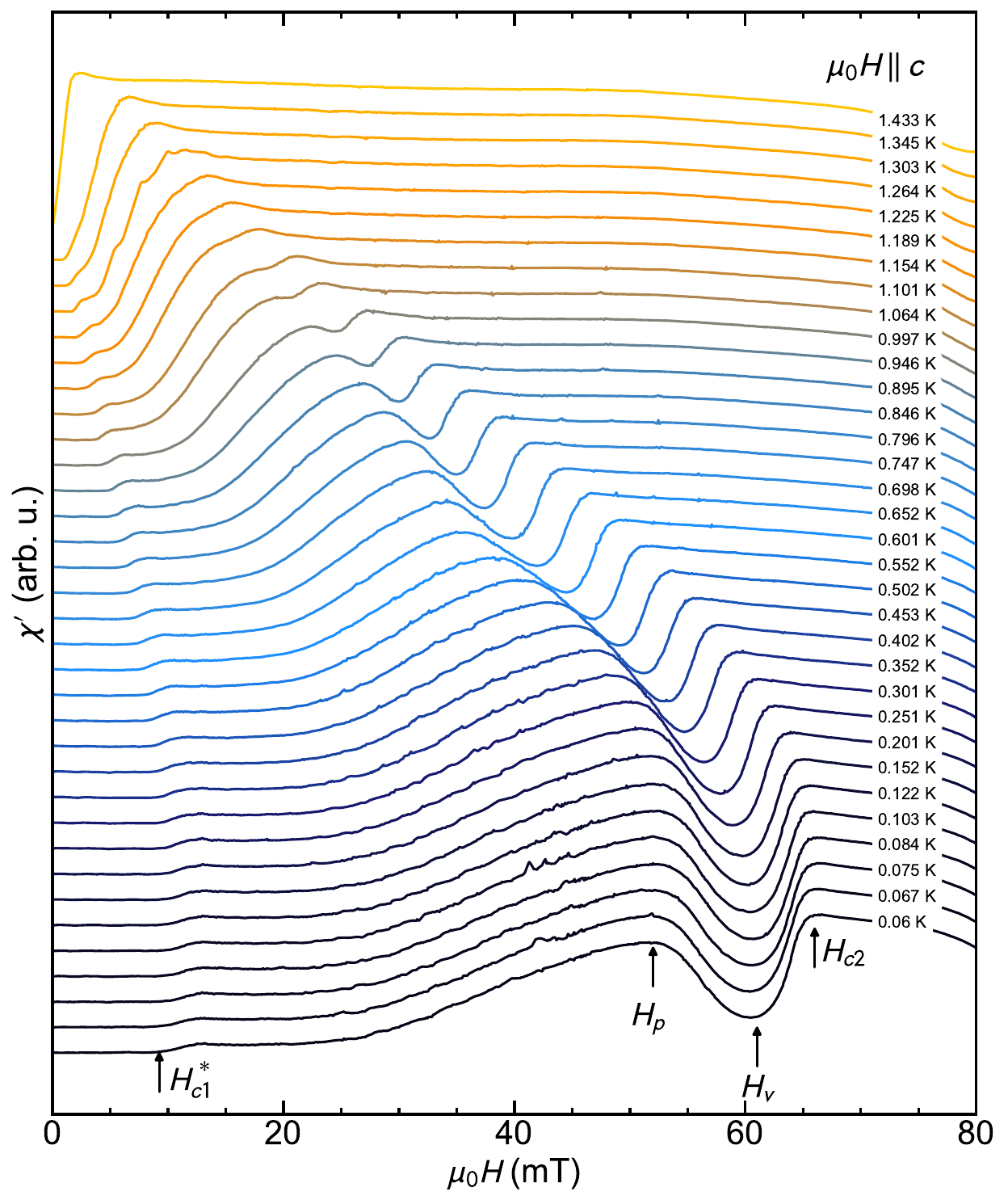}
	\caption{The complete data set of the magnetic field dependence of $\chi'(H)$ measurements at different temperatures. The arrows indicate the feature where $H^*_{c1}$,$H_p$,$H_p$ and $H_{c2}$ are defined.}
\label{fig:Figure_s4_raw_data}
\end{figure}

Fig. \ref{fig:Figure_s4_raw_data} shows part of the field sweeps (only up sweeps and positive fields) at all measured temperatures used to determine the phase diagram shown in the main paper. These curves are already shifted according to the procedure described above.


\subsection{The lower critical field and demagnetization factors}

\begin{figure}[t]
	\centering
	\includegraphics[width=\linewidth]{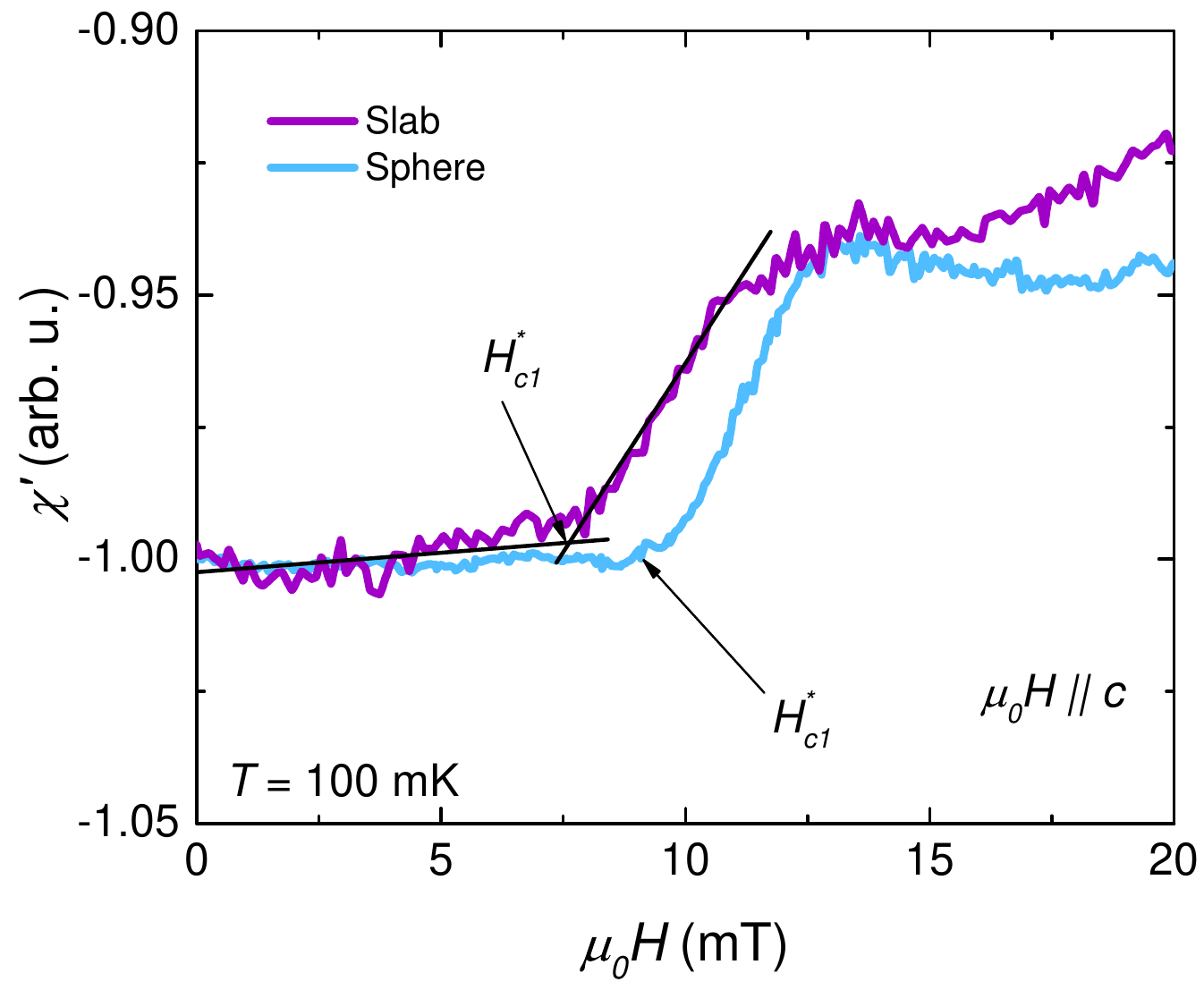}
	\caption{Comparison of $\chi'(H)$ near $H_{c1}$ at 100 mK between a spherical sample and slab-geometry sample.}
\label{fig:Figure_s5_comp}
\end{figure}

Figure \ref{fig:Figure_s5_comp} compares the measurement of $\chi'(H)$ near $H_{c1}$ of samples with a spherical and slab geometry. The measurements were done under the same conditions of 5\,Hz, 175\,$\mu$T excitation field, and 100\,mK. The feature in $\chi'(H)$ related to the $H_{c1}^*$ 
depends strongly on the geometry of the sample. 
From this measurement we obtain that $H^*_{c1,slab}\approx 7.5$ mT and $H^*_{c1,sphere}\approx 9.2$\,mT. If we take into account the demagnetization factor $N$:
\begin{equation}
    H_{c1}=\frac{H^*_{c1}}{N-1}
\end{equation}

The demagnetization factor for a sphere is $N=1/3$ and for a slab is described by \cite{Prozorov2018}:
\begin{equation}
    N\approx\frac{4ab}{4ab+3c(a+b)}
\end{equation} 

The dimensions $(a\times b\times c)$ of our slab sample are approximately $(500\times500\times 330)\,\mu$m leading to $N\approx0.5$. Then, $H_{c1,sphere}=13.8$\,mT and $H_{c1,slab}=15$\,mT. Both values are in the same order, within the experimental uncertainty on the estimation of $N$.

\subsection{Thermodynamic critical field $H_c$}
For a consistency check, we compare the thermodynamic critical field extracted from specific heat with the one extracted from our values of $H_{c1}$ and $H_{c2}$.
$H_c$ is determined from the free-energy difference between the normal and the superconducting states, $\Delta F= F_n-F_s$, obtained from the $C/T$ data by the integration of entropy difference \cite{Landaeta2022a}:

\begin{equation}
  \Delta F(T)=\frac{\mu_0H^2_c(T)}{2}=\int_{T_c}^{T}\int_{T_c}^{T'}\frac{C_s-C_n}{T''}dT''dT'.
\label{eq:Hc}
\end{equation}

\begin{figure}[t]
	\centering
	\includegraphics[width=\linewidth]{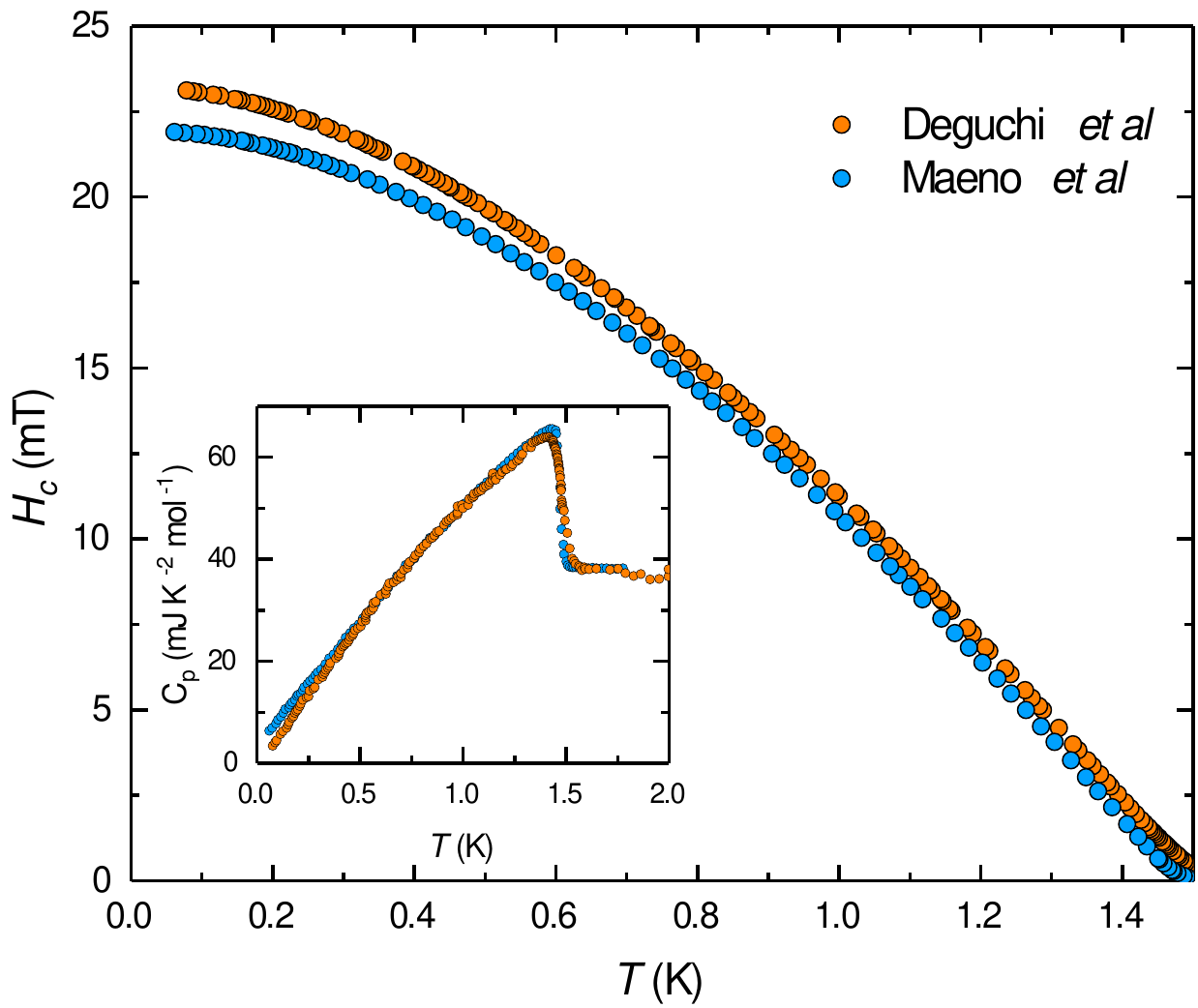}
	\caption{ Thermodynamic critical field $H_c$ estimated from specific heat measurement from Deguchi \textit{et. al.} \cite{Deguchi2004} and Maeno \textit{et. al.} \cite{MAENO2000}}
\label{fig:Figure_s6_hc}
\end{figure}

Here, $C_n$ is the value of $C/T$ at the normal state; this value varies from 38 to 40 mJK$^2$mol$^-1$, and $C_s$ is the $C_p$ measurement shown in the inset of Figure \ref{fig:Figure_s6_hc}. If we extrapolated $H_c$ to 0 K, it is conservative to say that $H_c(0)=(23\pm2)$ mT, which is consistent with previous reports \cite{Mackenzie2003}.

We can also calculate $H_c(0)$ using our measured values for $H_{c1}$ and $H_{c2}$.  Using $\frac{H_{c1}}{H_{c2}}=\frac{\ln{\kappa_\mathrm{GL}}+\alpha(\kappa_\mathrm{GL})}{2\kappa_\mathrm{GL}^2}$, where $\alpha(\kappa)$ is given in the Ref. \cite{Brandt2003}, we determine $\kappa_{\mathrm{GL}}=1.92$. This leads immediately to $H_c(0)=\frac{H_{c2}}{\sqrt{2}\kappa_\mathrm{GL}}=24.67$\,mT which is consistent, within experimental error, with the estimate from the analysis of the specific heat. 

\subsection{Coherence length and penetration depth from the critical fields}

The upper critical field is related to the superconducting coherence
length by 
\[
H_{c2}=\frac{\Phi_{0}}{2\pi\xi^{2}}
\]
On the other hand, the lower critical field is defined as 
\begin{equation}
H_{c1}=\frac{4\pi\epsilon_{1}}{\Phi_{0}}
\end{equation}
where $\epsilon_{1}$ is the free energy per unit length of a single
vortex line. To determine $\epsilon_{1}$ requires the solution of
a non-linear equation for the order parameter, while $H_{c2}$ is
determined by the linearized Ginzburg-Landau equations in the field.
A comprehensive
analysis of the linear and nonlinear Ginzburg-Landau theory was performed
by Brandt who finds 
\begin{equation}
H_{c1}=\frac{\Phi_{0}}{4\pi\lambda^{2}}C\left(\frac{\lambda}{\xi}\right)
\end{equation}
with $ C\left(\kappa\right)=\log\kappa+\alpha\left(\kappa\right)$.
The function $\alpha\left(\kappa\right)$ was determined numerically
in Ref. \cite{Brandt2003} for all values $\kappa>\kappa_{c}=\frac{1}{\sqrt{2}}$.
An analytic expression, that reproduces the numerical finding with
an accuracy better than $10^{-3}$, was also given 
 \cite{Brandt2003}
:
\begin{equation}
\alpha\left(\kappa\right)=\alpha_{\infty}+\exp\left(-c_{0}-c_{1}\log\kappa-c_{2}\left(\log\kappa\right)^{2}\right),
\end{equation}
with $\alpha_{\infty}=0.49693$, $c_{0}=0.41477$, $c_{1}=0.775$,
and $c_{2}=0.1303$. This expression  reproduces $C\left(\kappa_{c}\right)\approx 1$
at the onset of type II superconductivity and $C\left(\kappa\gg1\right)\approx\log\kappa+0.49693$,
as determined in Ref. \cite{Hu1972}. From the two critical fields, we
first construct

\begin{equation}
h_{c1}\equiv\frac{H_{c1}}{H_{c2}}=\frac{1}{2\kappa^{2}}\left(\log\kappa+\alpha\left(\kappa\right)\right)
\end{equation}
and determine $\kappa$  and $\xi$ using the
expression for $H_{c2}$ and then determine $\lambda=\kappa \xi$.

\subsection{Formalism of the nonlocal electrodynamics}

\subsubsection{Horizontal and vertical line nodes}

We consider gap functions on the Fermi surface that give rise to vertical
or horizontal line nodes, respectively. In the former case, we consider
$\Delta\left(\boldsymbol{k}\right)=\Delta_{0}\left(\cos k_{x}-\cos k_{y}\right)$.
In the low-energy limit we can safely write $\Delta\left(\varphi\right)=\Delta_{0}\cos2\varphi$
instead, where $\varphi$ is the polar angle of $\boldsymbol{k}$
in cylindrical coordinates. For the horizontal line nodes, we have
in mind a pairing state like $\Delta\left(\boldsymbol{k}\right)=\Delta_{0}\left(\sin k_{x}\pm i\sin k_{y}\right)\sin k_{z}$.
For our analysis only the magnitude of the gap $\left|\Delta\left(\boldsymbol{k}\right)\right|^{2}$ matters
and we use $\left|\Delta\left(\boldsymbol{k}\right)\right|^{2}=\Delta_{0}^{2}\sin^{2}k_{z}$,
i.e. we ignore the weak dependence of the gap amplitude on $k_{x,y}$.
The density of states 
\begin{equation}
\rho\left(\omega\right)=\int_{BZ}\frac{d^{3}k}{\left(2\pi\right)^{2}}\delta\left(\omega-\sqrt{\epsilon_{\boldsymbol{k}}^{2}+\Delta_{\boldsymbol{k}}^{2}}\right)
\end{equation}
 of both pairing states is the same and given as 
\begin{equation}
\rho\left(\omega\right)=\frac{2\rho_{F}}{\pi}{\rm Re}\left(K\left(\frac{\Delta_{0}^{2}}{\omega^{2}}\right)\right),
\end{equation}
where $K\left(x\right)$ is the elliptic integral of first kind. It
holds $\rho\left(\omega\ll\Delta_{0}\right)=\rho_{F}\omega/\Delta_{0}$,
$\rho\left(\omega\gg\Delta_{0}\right)=\rho_{F}$ and $\rho\left(\omega\approx\Delta_{0}\right)=\frac{\rho_{F}}{\pi}\log\frac{8\Delta_{0}}{\left|\omega-\Delta_{0}\right|}$.

In the local, London limit, the penetration depth is \cite{Scalapino1995} 

\begin{equation}
\lambda_{{\rm L}}^{-2}\left(T\right)=\lambda_{0}^{-2}\left(1+\delta k\left(T\right)\right)
\label{eq_lock}
\end{equation}
with

\begin{eqnarray}
\delta k\left(T\right) & = & \frac{2}{\rho_{F}}\int_{0}^{\infty}d\omega\rho\left(\omega\right)\frac{df\left(\omega\right)}{d\omega}.
\end{eqnarray}
It follows 
\begin{equation}
\left.\frac{\Delta\lambda\left(T\right)}{\lambda_{0}}\right|_{{\rm loc}}\equiv\frac{\lambda_{L}\left(T\right)-\lambda_{0}}{\lambda_{0}}=\frac{1}{\sqrt{1+\delta k\left(T\right)}}-1.\label{eq:loc_full}
\end{equation}
At small $T$ holds $\rho\left(\omega\right)=\rho_{F}\frac{\omega}{\Delta_{0}}$
and one obtains $\delta k\left(T\ll\Delta_{0}\right)=-2\log2\frac{T}{\Delta_{0}}$
which yields the celebrated result of a nodal superconductor $\left.\frac{\Delta\lambda\left(T\ll\Delta_{0}\right)}{\lambda_{0}}\right|_{{\rm loc}}=\log2\frac{T}{\Delta_{0}}.$
In our numerical analysis, we will use the full expression Eq.\ref{eq:loc_full}.
\[
\]

\subsubsection{Non-local electrodynamics}

The nonlocal relation between a current $\boldsymbol{j}$ and a vector
potential $\boldsymbol{A}$ (in London gauge: $\nabla\cdot\boldsymbol{A}=0$
together with $\boldsymbol{A}=0$ deep in the bulk and $\boldsymbol{A}\cdot\boldsymbol{n}=0$
at a surface with normal vector $\boldsymbol{n}$) is 
\begin{equation}
j_{\alpha}\left(\boldsymbol{q},\omega\right)=-K_{\alpha\beta}\left(\boldsymbol{q},\omega\right)A_{\beta}\left(\boldsymbol{q},\omega\right).
\end{equation}
We consider a superconductor in the half space $y>0$ and consider
the decay of the electromagnetic field along the $y$-direction. For
a magnetic field $\boldsymbol{B}=\nabla\times\boldsymbol{A}$ applied
along the $z$-direction we consider currents along the $x$-direction.
If we combine this relation with Maxwell's equation $\nabla\times\boldsymbol{B}=\frac{4\pi}{c}\boldsymbol{j}$
we get at $\omega=0$
\begin{equation}
A''_{x}\left(y\right)=-\frac{4\pi}{c}\int dy'K_{xx}\left(y-y'\right)A_{x}\left(y'\right).
\end{equation}
The penetration depth is commonly defined as
\begin{equation}
\lambda\left(T\right)=\frac{1}{B_{z}\left(y=0\right)}\int_{0}^{\infty}B\left(y\right)dy=-\frac{A_{x}\left(y=0\right)}{A'_{x}\left(y=0\right)},
\end{equation}
since $B_{z}\left(y\right)=A'_{x}\left(y\right)$, where the primes
are derivatives with respect to $y$. For specular boundary conditions
one finds after Fourier transformation \cite{Tinkham1974}
\begin{equation}
A_{x}\left(q_{y}\right)=\frac{B_{z}\left(y=0\right)}{q^{2}+\frac{4\pi}{c}K_{xx}\left(q_{y}\right)}
\end{equation}
Hence, it follows 
\begin{equation}
\lambda\left(T\right)=\frac{2}{\pi}\int_{0}^{\infty}\frac{dq_{y}}{q_{y}^{2}+\frac{4\pi}{c}K_{xx}\left(q_{y},T\right)}.\label{eq:pendepth_int}.
\end{equation}
The London penetration depth is  defined as 
\begin{equation}
  \lambda_{L}^{-2}=\frac{4\pi}{c}K_{xx}\left(q_{y}\rightarrow0\right).  
\end{equation}
If the transverse current kernel $K_{xx}\left(q_{y}\right)$ depends
weakly on momentum it is safe to assume $K_{xx}\left(q_{y}\right)\approx\frac{c}{4\pi}\lambda_{L}^{-2}$
and performing the integral yields $\lambda=\lambda_{L}$ as required.
The other crucial length scale of the superconductor is
the coherence length 
\begin{equation}
\xi_{0}=\frac{v_{F}}{\pi\Delta_{0}},
\end{equation}
 with Fermi velocity $v_{F}$ and gap amplitude $\Delta_{0}$. Nonlocal
effects go back to Pippard \cite{Pippard1953} who's kernel in momentum
space is discussed in Ref. \cite{Tinkham1974}. The key observation
is that $K_{xx}\left(q_{y}\gg\xi_{0}^{-1}\right)\approx\frac{3c}{16}\lambda_{L}^{-2}\left(q_t\xi_{0}\right)^{-1}$
decays for increasing momentum, i.e. magnetic field configurations with wave length smaller
than the coherence length induce  smaller screening currents.
Indeed, in the limit $\lambda_{L}\ll\xi_{0}$ the integral Eq.\ref{eq:pendepth_int}
is dominated by momenta $q_{y}>\xi_{0}^{-1}$ and yields the well known result of clean type I superconductors
$\lambda\sim\lambda_{L}\left(\xi_{0}/\lambda_{L}\right)^{1/3}$. 

Kosztin and Leggett \cite{Kosztin1997} found that non-local effects
are also important in nodal superconductors, yet in a more subtle
way, with different behavior of the kernel $K_{xx}\left(q_{y},T\right)$
for zero and finite temperatures. At $T=0$ the response is essentially
momentum independent and one can write 
\begin{equation}
K_{xx}\left(q_{y},0\right)=\frac{c}{4\pi}\lambda_{L}^{-2}\left(T=0\right).
\end{equation}
 Hence, for $T=0$ the actual penetration
depth is the London length and the electromagnetic response is local. For convenience we use
 $\lambda_{0}\equiv \lambda\left(T=0\right)$.  The non-locality
emerges at finite temperatures where the electromagnetic kernel
can be written as \cite{Kosztin1997}
\begin{eqnarray}
\frac{4\pi}{c}K_{xx}\left(q_{y},T\right) 
 & = & \lambda_{0}^{-2}+\left(\lambda_{{\rm L}}^{-2}\left(T\right)-\lambda_{0}^{-2}\right)F\left(\frac{\xi_T q_{y}}{\pi }\right).
 \label{Eq_kernal}
\end{eqnarray}
The characteristic length scale for the nonlocal  response due to nodes is the thermal de Broghlie wave length of massless Bogoliubov particles $\xi_T=v_F/(k_B T)$.
For vertical line nodes, the function $F\left(z\right)$ is given
as 
\begin{equation}
F_{{\rm v}}\left(z\right)=1-\frac{\pi z}{\sqrt{8}\log2}\int_{0}^{1}dyf\left(\frac{\pi}{\sqrt{8}}yz\right)\sqrt{1-y^{2}}.\label{eq:Fv}
\end{equation}
As small $z$, i.e. in the local limit holds $F_{{\rm v}}\left(z\ll1\right)=1-\frac{\pi^{2}z}{16\sqrt{2}\log2}$
while $F_{{\rm v}}\left(z\gg1\right)=\frac{6\zeta\left(3\right)}{\pi^{2}\log2}z^{-2}$.
The analogous result for horizontal line nodes is 
\begin{eqnarray}
F_{{\rm h}}\left(z\right) & = & 1-\frac{\pi z}{2\log2}\int_{0}^{1}dyf\left(\frac{\pi}{2}yz\right)\left(1-y\right)^{2}
\label{eq:Fh}
\end{eqnarray}
Now holds $F_{{\rm h}}\left(z\ll1\right)=1-\frac{\pi z}{12\log2}$
and $F_{{\rm h}}\left(z\gg1\right)=\frac{\pi}{3\log2}z^{-1}$.
The key difference between
horizontal and vertical line nodes stems from the distinct behavior at large $z$, i.e. for $ \xi_T q_y\gg 1$is, where  $F_{h}\left(z\gg1\right)\propto z^{-1}$
while $F_{v}\left(z\gg1\right)\propto z^{-2}$, respectively. Hence, the suppression
of non-local currents by thermal excitations is stronger for vertical
than for horizontal nodes.

In order to gain qualitative insights, we first analyze the leading low-temperature regime, where nonlocal corrections are small. For $T\ll\Delta_{0}$ follows
\begin{eqnarray}
\frac{\Delta\lambda\left(T\right)}{\lambda_{0}} 
 & = & \left.\frac{\Delta\lambda\left(T\right)}{\lambda_{0}}\right|_{{\rm loc}}\times \frac{4}{\pi}\int_{0}^{\infty}dq_{y}\lambda_{0}\frac{F\left(\frac{1}{\pi} \xi_T q_{y}\right)}{\left(1+\lambda_{0}^{2}q_{y}^{2}\right)^{2}},
\end{eqnarray}
where the first term describes the $T$-dependence in the  local approximation and the second term is the important correction. In the regime where $\xi_{T}\gg\lambda_{0}$ the penetration
depth is dominated by $F\left(z\gg1\right)$.  In the case  of vertical nodes holds
\begin{eqnarray}
\frac{\Delta\lambda\left(T\right)}{\lambda_{0}} & \sim & \left.\frac{\Delta\lambda\left(T\right)}{\lambda_{0}}\right|_{{\rm loc}}\int_{\xi_{T}^{-1}}^{\infty}dq_{y}\lambda_{0}\frac{1}{\left(\xi_{T}q_{y}\right)^{2}}
\nonumber \\
 & \sim & 
\left.\frac{\Delta\lambda\left(T\right)}{\lambda_{0}}\right|_{{\rm loc}}\frac{\lambda_{0}}{\xi_{T}},
\end{eqnarray}
where we dropped factors of order unity. This is the result given
in the main text that yields $\Delta\lambda\left(T\right)/\lambda_{0}\propto\kappa\frac{T^{2}}{\Delta_{0}^{2}}$.
The situation changes for horizontal nodes where 
\begin{eqnarray}
\frac{\Delta\lambda\left(T\right)}{\lambda_{0}} & \sim & \left.\frac{\Delta\lambda\left(T\right)}{\lambda_{0}}\right|_{{\rm loc}}\int_{\xi_{T}^{-1}}^{\lambda_{0}^{-1}}dq_{y}\lambda_{0}\frac{1}{\xi_{T}q_{y}}\nonumber \\
 & \sim & \left.\frac{\Delta\lambda\left(T\right)}{\lambda_{0}}\right|_{{\rm loc}}\frac{\lambda_{0}}{\xi_{T}}\log\frac{\xi_{T}}{\lambda_{0}},
\end{eqnarray}
which yields $\Delta\lambda\left(T\right)/\lambda_{0}\propto\kappa\frac{T^{2}}{\Delta_{0}^{2}}\log\frac{\Delta_{0}}{\kappa T}.$
These results  allow for a qualitative understanding of the impact of nodal excitations on the electromagnetic response. 
In our numerical analysis we avoided the expansion for small corrections
to the zero-temperature limit and used instead
\begin{eqnarray}
\frac{\Delta\lambda\left(T\right)}{\lambda_{0}}&=&-\frac{2\lambda_{0}}{\pi}\int_{0}^{\infty}dq_{y}\frac{\delta k\left(T\right)F\left(\frac{1}{\pi}\frac{v_{F}q_{y}}{k_{B}T}\right)}{\left(1+\lambda_{0}^{2}q_{y}^{2}\right)} \nonumber \\
&\times & \frac{1}{\left(1+\lambda_{0}^{2}q_{y}^{2}+\delta k\left(T\right)F\left(\frac{1}{\pi}\frac{v_{F}q_{y}}{k_{B}T}\right)\right)}. \label{eq_nonl_fin}
\end{eqnarray}
which follows from Eq.\eqref{Eq_kernal} together with Eq.\eqref{eq_lock}.
Similar expressions can also be obtained for diffuse boundaries.

\begin{figure}[htbp]
	\centering
	\includegraphics[width=\linewidth]{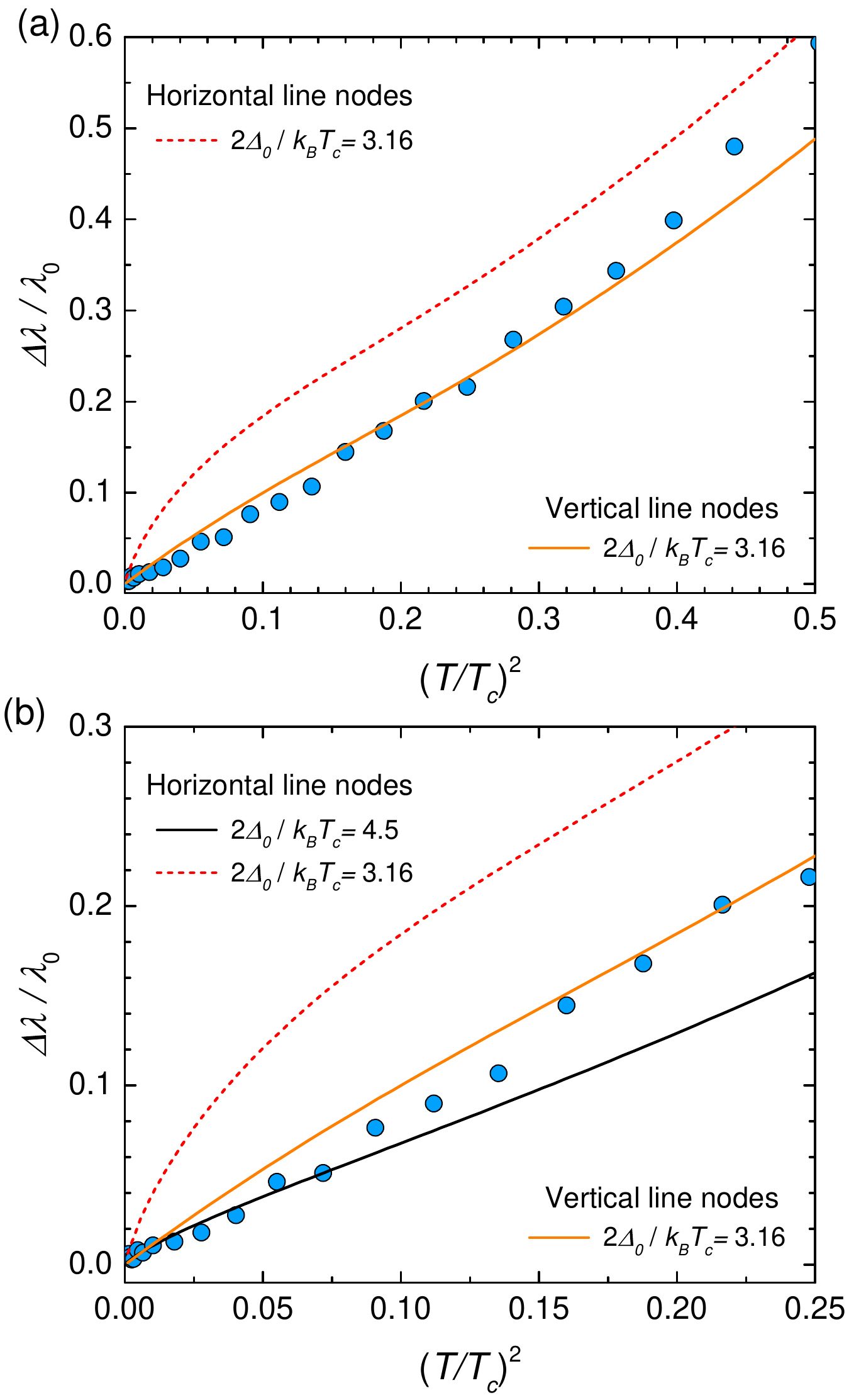}
	\caption{(a) $\Delta\lambda\left(T\right)/\lambda_0$ vs $(T/T_c)^2$, the red dashed and orange solid lines are the expected behavior of horizontal and vertical line nodes respectably in the non-local limit with $2\Delta_0/k_B T_c= 3.16$. (b) The black line is the best fit for horizontal line nodes with $2\Delta_0/k_B T_c= 4.5$}
\label{fig:Figure_s7_fits}
\end{figure}
The dimensionless $T$-variation of the penetration depth   $\Delta\lambda\left(T\right)/\lambda_0$, plottet  as function of the dimensionless temperature variable $T/T_c$,  only depends on the $T=0$ ratio $\kappa_0=\lambda_0/\xi_0$ and on $2\Delta_0/(k_B T_c)$. The former is determined by our simultaneous measurement of $H_{c1}$ and $H_{c2}$. For the latter we used the BCS result of $3.53$ in the main text. In our analysis, the gap amplitude $\Delta_0$ determines the slope of the gap near the node which does not have to agree with the overall gap maximum, as determined by spectroscopic means, in particular for multiband superconductors. This slope also determines the low-$T$ dependence of the specific heat $C_{s}=\tfrac{9\zeta\left(3\right)\rho_{F}}{2}\tfrac{T^{2}}{\Delta_{0}}$, where we used the low-frequency behavior of the density of states, given earlier. A $T$-dependence $C_{s}=\alpha T^{2}$ of the heat capacity was indeed observed in Ref.~\cite{NishiZaki2000} where $\alpha \approx 52.7\:{\rm mJ/(mol\:K}^{3})$, while in the normal state $C_{n}=\gamma T$ with $\gamma=37.5\:{\rm mJ}/\left({\rm mol\,K}^{2}\right)$. 
Hence, the low-energy slope follows as $\Delta_{0}=\tfrac{27\zeta\left(3\right)}{\pi^{2}}\tfrac{\gamma}{\alpha}$ which yields $2\Delta_0/(k_B T_c)=3.16$. In Fig.~\ref{fig:Figure_s7_fits}a we show the penetration depth due to nonlocal excitations for this value of the gap for horizontal and vertical line nodes, respectively.
The agreement between experiment and theory for vertical line nodes is even slightly better for this value of the gap amplitude.
Finally, given the dependence of our results on $2\Delta_0/(k_B T_c)$ we also plot in Fig.~\ref{fig:Figure_s7_fits}b our best fit of the data to the theory with horizontal line nodes (black curve). Clearly, even if we allow the gap amplitude to vary as fit parameter, we cannot achieve reasonable agreement between horizontal line nodes and our data.


%

\end{document}